\documentclass[11pt]{article}
\usepackage{amsmath,amssymb,amsfonts}

\newtheorem{thm}{Theorem}

\newtheorem{lem}{Lemma}
\def\rem{{\noindent{\bf Remark.}\ }}

\def\sds{\strut \displaystyle}

\title{Fast Arithmetics Using Chinese Remaindering}
\author{George Davida\thanks{Department of EE \& CS,
University of Wisconsin-Milwaukee, WI, USA; e-mail: {\tt
davida@cs.uwm.edu}}, Bruce Litow\thanks{School of Information
Technology, James Cook University, Townsville, QLD, Australia;
e-mail: {\tt bruce@cs.jcu.edu.au}} and Guangwu Xu\thanks{Department
of EE \& CS, University of Wisconsin-Milwaukee, WI, USA; e-mail:
{\tt gxu4uwm@uwm.edu}}}

\date{}

\begin{document}
\maketitle

\begin{abstract}
In this paper, some issues concerning the Chinese remaindering
representation are discussed. Some new converting methods, including
an efficient probabilistic algorithm based on a recent result of von
zur Gathen and Shparlinski \cite{Gathen-Shparlinski}, are described.
An efficient refinement of the NC$^1$ division algorithm of Chiu,
Davida and Litow \cite{Chiu-Davida-Litow} is given, where the number
of moduli is reduced by a factor of $\log n$.

\

\end{abstract}

\section{Introduction}

For the fundamental arithmetic operations, it is often desirable to
represent an integer as a vector of smaller integers. This can be
done by selecting a set of pairwise coprime positive integers $m_1,
m_2, \dots, m_r$, and mapping an integer $x$ to the vector of
residues $(|x|_{m_1}, |x|_{m_2},\cdots, |x|_{m_r})$, where
$|x|_{m_i}$ denotes $x \pmod {m_i}$. This approach is called the
{\sl Chinese remaindering representation (CRR)}, as the {\sl Chinese
remainder theorem (CRT)} guarantees such mapping is meaningful.
Using CRR, large calculations can be split as a series of smaller
calculations that can be performed independently and in parallel.
So, this approach has a significant role to play in applications
such as cryptography and high precision scientific computation.

It is well known that three basic arithmetic operations, addition,
subtraction, and multiplication, can be performed in $O(\log n)$
time using $n^{O(1)}$ processors. These operations can also be done
in the manner of log-space uniform. However, the parallel complexity
of integer division is a subtle problem and has attracted a lot of
attention. The first $O(\log n)$ time $n^{O(1)}$ sized circuit for
integer division was exhibited by Beame, Cook and Hoover
\cite{Beame-Cook-Hoover}. Recently, the log-depth, polynomial size,
logspace-uniform circuit family for integer division (i.e., integer
division is in logspace-uniform NC$^1$) was described by Chiu,
Davida and Litow \cite{Chiu-Davida-Litow}. This settled a
longstanding open problem and provided an optimal computation
efficiency theoretically.

In this paper, we discuss some issues concerning the Chinese
remaindering representation. The organization of the paper is as
follows. Section 2 describes the Chinese remaindering system. Two
methods for converting a vector to the corresponding
integer are presented in this section. Section 3 focuses on the
integer division using CRR. Under the framework of NC$^1$, an
efficient refinement of the division algorithm of Chiu, Davida and
Litow \cite{Chiu-Davida-Litow} is proposed.

\section{Chinese Remainder Representation}

Let ${\cal M}=\{m_1, m_2, \dots , m_r\}$ be a set of pairwise
coprime integers and $\sds M=\prod_{i=1}^r m_i$. For a set of
integers $x_1, x_2, \dots, x_r$ with $0\le x_i < m_i$, the Chinese
Remainder Theorem says that the system of congruence
\[
\left\{ \begin{array}{l} x\equiv x_1 \pmod {m_1} \\
                               x\equiv x_2 \pmod {m_2} \\
                                \cdots \\
                               x\equiv x_r \pmod {m_r}
\end{array} \right.
\]
has a unique solution $0\le x < M$. In fact, using the extended
Euclidean algorithm, one finds integers $u_1, u_2, \cdots, u_r$ such
that
\[
\sum_{i=1}^r u_i \frac{M}{m_i} = 1,
\]
and it is easy to verify that
\begin{equation}\label{eq:2.1}
x = \sum_{i=1}^r x_i u_i \frac{M}{m_i} \pmod M
\end{equation}
gives the desired solution. It is remarked that one can also choose
$\sds u_i=( \frac{M}{m_i} )^{-1}\pmod{m_i}$; and such choice of
$u_i$ will be used in the rest of our discussion.

The above system is called a Chinese remaindering representation
(CRR) based on the set ${\cal M}$, and is denoted by CRR(${\cal
M}$).

Now we present a method of finding $u_i$'s which can be seen as an
alternative to the Garner algorithm described in \cite{knuth} (pages
290,293).

For each $j>1$, $m_j$ is coprime to $m_1\cdots m_{j-1}$. Therefore,
by the extended Euclidean algorithm, there exist integers $\alpha_j,
\beta_j$ such that
\begin{equation}\label{eq:2.2}
\alpha_j m_j + \beta_j m_1\cdots m_{j-1} = 1.
\end{equation}
With these $r-1$ pairs of $(\alpha_i, \beta_i)$, the coefficients
$u_i$ can be computed as follows:
\begin{eqnarray*}
u_1&\gets&\alpha_2\alpha_3\cdots \alpha_r  \pmod{m_1}\\
u_2&\gets&\beta_2\alpha_3\cdots \alpha_r  \pmod{m_2}\\
u_3&\gets&\beta_3\alpha_4\cdots \alpha_r  \pmod{m_3}\\
& & \dots  \\
u_r & \gets & \beta_r  \pmod{m_r}
\end{eqnarray*}

The correctness of the above algorithm is based on the following
identity:
\begin{eqnarray*}
&&(\alpha_2\cdots \alpha_r) m_2m_3\cdots m_r +
(\beta_2\alpha_3\cdots
\alpha_r) m_1m_3\cdots m_r +\\
&& (\beta_3\alpha_4\cdots \alpha_r) m_1m_2m_4\cdots m_r+\cdots
+\beta_r m_1m_2\cdots m_{r-1}=1.
\end{eqnarray*}

This identity can be verified using the standard mathematical
induction: for $i> 2$, suppose that
\begin{eqnarray*}
&&(\alpha_2\cdots \alpha_{i-1}) m_2m_3\cdots m_{i-1} +
(\beta_2\alpha_3\cdots
\alpha_{i-1}) m_1m_3\cdots m_{i-1} +\\
&& (\beta_3\alpha_4\cdots \alpha_{i-1}) m_1m_2m_4\cdots
m_{i-1}+\cdots +\beta_{i-1} m_1m_2\cdots m_{i-2}=1.
\end{eqnarray*}
Multiply both sides of the above by $\alpha_im_i$, and apply the
equation (\ref{eq:2.1}) for $j=i$, one gets
\begin{eqnarray*}
&&(\alpha_2\cdots \alpha_i) m_2m_3\cdots m_i +
(\beta_2\alpha_3\cdots
\alpha_i) m_1m_3\cdots m_i +\\
&& (\beta_3\alpha_4\cdots \alpha_i) m_1m_2m_4\cdots m_i+\cdots
+\beta_i m_1m_2\cdots m_{i-1}=1.
\end{eqnarray*}

It is remarked that in this process, we call the extended Euclidean
algorithm $r-1$ times. For the method described in \cite{knuth},
$\sds \frac{r(r-1)}2$ instances of extended Euclidean algorithm need
to be invoked, for pairs $(m_i, m_j)$ with $i<j$.

\

Next we present a probabilistic converting method for CRT. For
positive integers $N_1, N_2$, let $a_1 , a_2 , \cdots , a_r$ be
 in $\{1,2,\cdots, N_1\}$. Pick $2r$ uniformly
distributed random integers $s_1 , s_2 , \cdots  , s_r$ and $t_1 ,
t_2 , \cdots , t_r$ in $\{1, 2, \cdots, N_2\}$ and consider the
linear forms
\[
S=\sum_{i=1}^r a_is_i, \mbox{  } T=\sum_{i=1}^r a_it_i.
\]
It has been proved by Cooperman, Feisel, von zur Gathen and Havasin
in \cite{CFGH} that with high probability
\begin{equation}\label{eq:2-3}
\gcd(a_1, a_2, \cdots, a_r) = \gcd(S, T).
\end{equation}
This was improved recently by von zur Gathen and Shparlinski
\cite{Gathen-Shparlinski} and they gave the following strong result:
with probability at least $\sds \frac{6}{\pi^2}+o(1)$,
\[
\gcd(a_1, a_2, \cdots, a_r) = \gcd(S, T),
\]
provided that $\sds \frac{N_2}{r+\ln N_1}$ is large enough.

This result can be used to produce a very efficient probabilistic
algorithm for Chinese remaindering. Let us take $a_i =
\frac{M}{m_i}$. We can find $x$ such that
\[
\left\{ \begin{array}{l} x\equiv x_1 \pmod {m_1} \\
                               x\equiv x_2 \pmod {m_2} \\
                                \cdots \\
                               x\equiv x_r \pmod {m_r}
\end{array} \right.
\]
by the following steps:
\begin{enumerate}
\item Choose random linear forms $S, T$ until
\[
\gcd(S, T) = 1.
\]
(The expected number for getting the desired pair of $S, T$ is less
than $2$.)
\item Use extended Euclidean algorithm to get integers $u, v$ such
that
\[
uS+vT = \sum_{i=1}^r (us_i+vt_i)\frac{M}{m_i} = 1.
\]
\item The solution $x$ is
\[
x = \sum_{i=1}^r x_i(us_i+vt_i)\frac{M}{m_i} \pmod {M}.
\]
\end{enumerate}

\rem It can be seen that in this routine, if the extended Euclidean
algorithm is used to compute all {\sl gcd}s, then the expected
number of rounds to get $u, v$ in step 2 is less than $2$. In step
3, $us_i+vt_i$ can be replaced by $(us_i+vt_i) \pmod {m_i}$.

\section{An Improved NC$^1$ Division Algorithm}

In this section, we discuss the division algorithm of Chiu, Davida
and Litow \cite{Chiu-Davida-Litow}. A careful analysis enables us to
reduce the number of prime moduli by a factor of $\log n$.

Let $\alpha$ be a real number. A rational number $\alpha'$ is said
to be an {\it $n-$bit under approximation to $\alpha$} if
\[
0\le \alpha - \alpha' \le \frac{1}{2^n}.
\]

The next result improves the lemma 3.2 of \cite{Chiu-Davida-Litow}:
\begin{lem}\label{lem:3.1}
Let $\frac{1}2 \le \alpha < 1$ and $\beta = 1-\alpha$. If $\sds
\frac{t_1}{A_1},\frac{t_2}{A_2},\dots,\frac{t_{n+1}}{A_{n+1}}$ are
$(n+3)-$bit underapproximations to $\beta$, then
\[
1+\frac{t_1}{A_1}+\frac{t_1t_2}{A_1A_2}+\cdots+\prod_{i=1}^{n+1}\frac{t_i}{A_i}
\]
is an $n-$bit underapproximation to $\sds \frac{1}{\alpha}$.
\end{lem}

{\it Proof}. Let
\[
\eta = \min_{1\le i \le n+1}\{\frac{t_i}{A_i}\}.
\]
Note that $\sds 0\le \beta \le \frac{1}2$ and $\sds 0\le \beta-\eta
\le \frac{1}{2^{n+3}}$, we see that

\begin{eqnarray*}
\frac{1}{\alpha}-(1+\frac{t_1}{A_1}+\frac{t_1t_2}{A_1A_2}+\cdots+\prod_{i=1}^{n+1}\frac{t_i}{A_i})
&\le & \frac{1}{\alpha}-(1+\eta+\eta^2+\cdots +\eta^{n+1})\\
&=&\frac{1}{1-\beta}-\frac{1-\eta^{n+2}}{1-\eta}\\
&=& \big( \frac{1}{1-\beta}-\frac{1}{1-\eta} \big)+\frac{\eta^{n+2}}{1-\eta}\\
&=&  \frac{\beta-\eta}{(1-\beta)(1-\eta)}+\frac{\eta^{n+2}}{1-\eta}\\
&\le& \frac{\frac{1}{2^{n+3}}}{\frac{1}2\cdot
\frac{1}2}+\frac{\frac{1}{2^{n+2}}}{\frac{1}2}\\
&=&\frac{1}{2^n}.\\
\end{eqnarray*}

In \cite{Chiu-Davida-Litow}, the log-depth, polynomial size,
logspace-uniform circuit family for integer division was constructed
by Chiu, Davida and Litow. In other words, integer division is
proved to be in logspace-uniform NC$^1$. This solves a longstanding
open problem.

Notice that the original construction of the NC$^1$ circuit family
for integer division needs $3n^2$ (actually $2n^2+5n$) primes
numbers. The main purpose of this section is to refine the
Chiu-Davida-Litow construction to achieve more efficiency. To be
more specific, we shall show that $\sds \frac{n^2}{\log n} +3n$
primes will be sufficient.

\begin{thm}
The number of prime moduli of the Chiu-Davida-Litow NC$^1$ integer
division algorithm can be reduced to $\sds \frac{n^2}{\log n}+3n$.
\end{thm}

{\it Proof}. The proof follows the similar line as in
\cite{Chiu-Davida-Litow}.

The goal is: given $x, y < 2^n$, compute the CRR of $\sds
\bigg\lfloor \frac{x}y \bigg\rfloor$.

Let $N=\sds \bigg\lfloor \frac{n^2}{\log n}\bigg\rfloor + 3n$.

Suppose that $x, y$ are represented in a CRR system with base
$\{m_1,m_2, \dots, m_n\}$ where $m_i$ is the $(i+2)$th prime
($m_1>3$). This base is extended to
\[
\{m_1,m_2, \dots, m_n, m_{n+1}, \dots, m_N\}.
\]

A product $D$ of the initial part of the base and some power of $2$
will be constructed so that
\[
\frac{1}2 \le \frac{y}D < 1.
\]
According to \cite{Chiu-Davida-Litow}, if $y = 2$, set $D=2$. If
$y>2$, then take $j<n$ to be the number such that
\[
m_1m_2\cdots m_j \le y < m_1m_2\cdots m_jm_{j+1}.
\]
Let $k$ be the smallest positive integer such that
$y<2^km_1m_2\cdots m_j$ (therefore $\sds \frac{y}{2^km_1m_2\cdots
m_j}\ge \frac{1}2$), and set
\[
D=2^km_1m_2\cdots m_j.
\]

Let $\sds r = \big\lfloor \frac{n}{\log n}\big \rfloor$. If $n\ge
2^6$, then $\sds \frac{n-\log n -(\log n)^2}{\log n} > 3$. The fact
that $m_{n+1} > 2n $  gives
\begin{eqnarray}\label{ineq:3.1}
(m_{n+1})^r &>& (2n)^{\big\lfloor \frac{n}{\log n}\big \rfloor} \nonumber \\
            &\ge & \big(2^{\log n +1}\big)^{\frac{n}{\log n}-1} \nonumber \\
            & = & 2^{n+\frac{n-\log n - (\log n)^2}{\log n}} \nonumber \\
            & > & 2^{n+3}
\end{eqnarray}
Since $n+(n+1)r \le N$, we can form the following products:
\begin{eqnarray*}
A_1 &=& m_{n+1}m_{n+2}\cdots m_{n+r}\\
A_2 &=& m_{n+r+1}m_{n+r+2}\cdots m_{n+2r}\\
&& \cdots\\
A_{n+1} &=& m_{n+nr+1}m_{n+nr+2}\cdots m_{n+(n+1)r}.\\
\end{eqnarray*}
We note that $A_i > 2^{n+3}$ for $i = 1, 2, \dots, n+1,$ by
(\ref{ineq:3.1}).

Next, choose
\[
t_i=\big\lfloor \frac{(D-y)A_i}{D}\big\rfloor, \mbox{ for } i = 1,
2, \dots, n+1.
\]
Similar to \cite{Chiu-Davida-Litow}, $\sds \frac{t_i}{A_i}$ can be
computed in NC$^1$. It is also routine to check that $\sds
\frac{t_i}{A_i}$ is an $(n+3)-$bit underapproximation to $\sds
\beta= \frac{D-y}D$.

Finally, by the lemma~\ref{lem:3.1}, we get an $n-$bit
underapproximation to $\sds \frac{1}{\alpha}$ where $\alpha = \sds
\frac{y}D$:
\[
\gamma = 1+\frac{t_1}{A_1}+\frac{t_1t_2}{A_1A_2}+\cdots
+\frac{t_1t_2\cdots t_{n+1}}{A_1A_2\cdots A_{n+1}}.
\]

Again, similar to \cite{Chiu-Davida-Litow}, we have
\[
\big\lfloor \frac{x}{y}\big\rfloor = \big\lfloor
x\frac{\gamma}{D}\big\rfloor \mbox{ or } \big\lfloor
\frac{x}{y}\big\rfloor = \big\lfloor x\frac{\gamma}{D}\big\rfloor+1.
\]
And all the computations are done in NC$^1$.

\rem The Chebyshev bounds for primes can be used  to get an
inequality which is a bit sharper than the inequality
(\ref{ineq:3.1}), but there is no significant reduction on the
number of prime moduli.

\end{document}